\documentclass[conference]{IEEEtran}

\IEEEoverridecommandlockouts
\usepackage{amsmath,amssymb,amsfonts}
\usepackage{algorithmic}
\usepackage{graphicx}
\usepackage{bmpsize}
\usepackage{xcolor}
\usepackage{lipsum}

\usepackage{graphicx} 
\usepackage[table]{xcolor}
\usepackage{multirow}
\usepackage{enumitem}
\usepackage{xspace}
\usepackage{caption}
\usepackage{svg}

\usepackage{cite}
\usepackage{amsmath,amssymb,amsfonts}
\usepackage{algorithmic}
\usepackage{graphicx}
\usepackage{textcomp}
\usepackage{bmpsize}
\usepackage{xcolor}
\usepackage{lipsum}

\newcommand{\ournameNoSpace}{\mbox{VocalBridge}}
\newcommand{\ourname}{\ournameNoSpace\xspace}

\usepackage{xurl}
\usepackage[colorlinks=true,urlcolor=black]{hyperref}
\def\BibTeX{{\rm B\kern-.05em{\sc i\kern-.025em b}\kern-.08em
    T\kern-.1667em\lower.7ex\hbox{E}\kern-.125emX}}
\begin{document}

\title{VocalBridge: Latent Diffusion-Bridge Purification for Defeating Perturbation-Based Voiceprint Defenses
}

\author{
    \IEEEauthorblockN{Maryam Abbasihafshejani}
    \IEEEauthorblockA{Department of Computer Science\\ University of Texas at San Antonio\\ San Antonio, Texas}
    \and
    \IEEEauthorblockN{AHM Nazmus Sakib}
    \IEEEauthorblockA{Department of Computer Science\\University of Texas at San Antonio\\ San Antonio, Texas}
    \and
    \IEEEauthorblockN{Murtuza Jadliwala}
    \IEEEauthorblockA{Department of Computer Science\\University of Texas at San Antonio\\ San Antonio, Texas}
}
\maketitle
\begin{abstract}
The rapid progress of speech synthesis technologies, such as text-to-speech (TTS) and voice conversion (VC), has intensified security and privacy concerns surrounding voice cloning. Recent defenses attempt to prevent unauthorized cloning by embedding protective perturbations into speech, aiming to obscure speaker identity while maintaining intelligibility. However, adversaries can employ state-of-the-art purification strategies to remove these perturbations, recover genuine acoustic characteristics, and regenerate cloneable voices. Although such threats are increasingly realistic, the robustness of these protective mechanisms under adaptive purification has not been adequately studied.

Existing purification and denoising approaches, mostly designed for adversarial noise in automatic speech recognition (ASR) or classification, fail to preserve the fine-grained acoustic features that define a speaker’s voice. As a result, they often degrade the perceptual quality and introduce distortions in the speaker-embedding space. 
Recent diffusion-based purification frameworks provide stronger denoising against perturbations targeting Automatic Speaker Verification (ASV) systems, but many of them operate directly on the waveform or require transcript-dependent phoneme alignment during inference. Although they outperform ASR-oriented denoising methods, these approaches still struggle to fully remove adversarial perturbations, which limits recovery fidelity and reduces overall performance.
 To address these limitations, we propose \emph{Diffusion-Bridge (\ourname)}, a purification model that learns a latent mapping from perturbed to clean speech within the EnCodec latent space. It employs a time-conditioned 1D-UNet denoiser to perform reverse diffusion under a cosine noise schedule, enabling efficient, transcript-free purification while preserving speaker-discriminative cues.
We also introduce a Whisper-Guided Phoneme variant that incorporates lightweight linguistic conditioning from a Whisper-based phoneme alignment module. Unlike prior text-conditioned diffusion models that rely on transcripts or external language prompts, our approach operates entirely in the acoustic domain and requires no linguistic supervision

These findings expose the fragility of current perturbation-based defenses and underscore the need for more resilient safeguards against evolving voice-cloning threats.
\end{abstract}
\section{Introduction}

Recent advances in generative AI have intensified concerns about security, privacy, and data ownership. A particularly troubling area is speech generation, including text-to-speech (TTS) and voice conversion (VC) models~\cite{ballesteros2021deep4snet}. These technologies can now produce highly realistic audio deepfakes that enable impersonation, misinformation, and identity theft~\cite{nici2024fcc,nyt2024scamworld,semansky2024fake}. For example, scammers recently cloned the voice of a public figure to deceive a family member into believing that they were in legal trouble~\cite{nypost2024shooster}. Critically, AI-generated fake voices can bypass state-of-the-art ASV, which still face challenges of generalization and robustness~\cite{jamdar2025syntheticpop}. Beyond individual fraud cases, synthetic speech has already been used in large-scale political misinformation campaigns, fraudulent financial transactions, and social engineering attacks targeting enterprises and government agencies. In February~2025, Italian authorities uncovered an AI-voice scam in which criminals impersonated the Italian Defence Minister and used voice-cloning technology to trick a prominent entrepreneur into wiring nearly~€1\,million ($\approx$ \$1.04\,million) to a foreign account\cite{reuters2025italyvoice}. In response to these escalating threats, both OpenAI~\cite{openai2024syntheticvoices} and the U.S.\ Federal Trade Commission (FTC)~\cite{ftc2024comment} have released reports warning about the security, privacy, and societal implications of voice conversion and synthetic voice technologies, highlighting the urgency of developing robust methods to detect and mitigate malicious uses of modern speech generation systems.

To mitigate these risks, researchers have begun exploring proactive voice protection methods that make a user’s speech \textit{unlearnable} for synthesis models. The key idea is to introduce carefully crafted perturbations or targeted transformations that prevent TTS/VC systems from extracting speaker-specific features (timbre), while keeping the audio natural and intelligible for legitimate use such as communication or authentication by ASV sytems. These perturbations reduce the ability of synthesis models to accurately mimic the identity of a speaker, ensuring that any fake audio generated from protected speech is likely to be of lower quality or unable to pass ASV systems
\cite{antifake,safespeech}.

However, the performance of existing proactive defenses remains underexplored in scenarios where attackers can purify protected speech before using it. Prior purification research has primarily focused on removing adversarial noise targeting automatic speech recognition (ASR) systems, leaving open the question of whether current methods can remove protective perturbations designed to defend against voice cloning and speaker verification attacks. To demonstrate the insufficiency of existing purification approaches and the need for more robust speech protection techniques, we introduce \emph{\ourname}, a comprehensive purification attack that effectively removes protective perturbations and enables attackers to bypass modern speaker verification systems.

\ourname is a \emph{diffusion-bridge} purification model that learns a latent mapping from protected to clean speech representations within a compact encoded domain. The model employs a time-conditioned 1D U-Net denoiser that models the reverse diffusion process inside the EnCodec latent space, trained to estimate residual perturbations under a cosine-based noise schedule. This design enables efficient, transcript-free purification while preserving the speaker characteristics needed for cloning and verification attacks. Our results show that \ourname substantially surpasses existing purification methods in both attack  effectiveness and performance 

Our contributions can be summarized as follows:

\begin{enumerate}
    \item \textbf{Comprehensive evaluation of proactive defenses.}
We conduct a large-scale evaluation of five state-of-the-art proactive defenses across six modern voice-synthesis tools and three widely used ASV encoders, providing an up-to-date view of how current protection mechanisms behave under diverse synthesis pipelines. Our study incorporates recent synthesis models, multiple defense strategies and purification tools, and a large dataset constructed from both VCTK and LibriSpeech, resulting in a comprehensive assessment of proactive protection in contemporary voice-cloning scenarios.
    \item \textbf{\ourname Diffusion-Bridge purification in the speech latent space.}
We introduce a bridge-diffusion purification model that operates in the speech latent space and learns a reverse mapping from diffused defensive perturbations representations back to their clean latent counterparts. We further enhance performance with a lightweight Whisper-based conditioning mechanism that provides phoneme-level guidance without requiring transcripts. Across our evaluation metrics, Verification Recovery (VR), Mean Opinion Score (MOS), and Word Error Rate (WER), \ourname consistently improves authentication restoration, perceptual quality, and intelligibility compared to recent purification methods, while also demonstrating strong generalization and robustness to adaptive protection strategies.
    \item \textbf{Dataset creation for community evaluation.}
    We generate a large dataset that includes clean, protected, and purified speech samples suitable for evaluating proactive defenses, providing a useful benchmark for future research.
\end{enumerate}

\section{Background and Related Work} 
\subsection{Deepfake Speech Synthesis Techniques}

Deepfake speech synthesis involves the use of AI and machine learning techniques to generate audio clips that replicate the vocal characteristics of a target speaker, including timbre, prosody, and articulation patterns. 
Deepfake speech synthesis techniques can be broadly classified into: (i) TTS and (ii) VC tools. 


TTS tools convert written text into spoken words. Some of the earliest techniques, such as concatenative synthesis, were capable of producing clear speech; however, the result often sounded unnatural and robotic \cite{Tabet2011SpeechST,noauthor_2011-qi}. In contrast to these, modern deep learning approaches can generate remarkably natural and human-like vocal output~\cite{sd61,sd62}. These systems typically operate in a two-stage pipeline. First, an acoustic model converts the input text into a low-level acoustic representation, such as a Mel spectrogram, which encodes phonetic information, pitch, and timing. Then, a neural vocoder synthesizes a high-fidelity audio waveform from this representation. The use of deep neural networks in both stages is what allows these models to capture the complex nuances of human speech. A key application of this technology is \emph{voice cloning}~\cite{sd17}, which aims to create a synthetic voice that sounds indistinguishable from a specific target speaker. This is achieved by training or fine-tuning a TTS model on a collection of audio recordings of that individual. By learning from these samples, the model captures the unique vocal characteristics of the target, including their timbre, intonation patterns, and accent~\cite{sd161,sd162,sd163}. This allows the trained model to be able to effectively generate a cloned voice.

Unlike TTS, voice conversion or VC modifies a source speaker's vocal attributes to match those of some target speaker, while preserving the original linguistic content. This is achieved by separating speaker-dependent acoustic features from the phonetic information and re-synthesizing the audio. 

\subsection{Speech Privacy and Protection Mechanisms}
Deepfake speech synthesis can seriously hamper user privacy and create security risks, consequently various defensive strategies have been explored in literature. Although anonymizing a speaker’s voice is an effective strategy to protect privacy~\cite{sd91,sd90}, it is not sufficient on its own as a defense; protection against deepfake audio also requires accurate detection and prevention. Detection focuses mainly on the identification of synthetic (or cloned) audio and includes techniques such as classification using Mel-frequency cepstral coefficient (MFCC) features~\cite{sd92}, emotion recognition~\cite{sd93}, acoustic signal analysis by simulating auditory effects of the human ear~\cite{sd94} and ASV systems~\cite{sd95}. Preventive defense techniques are more retroactive in nature and aimed at rendering the speech synthesizing capability of audio synthesis tools ineffective. To combat the creation of fake or synthetic speech, several techniques have been proposed that employ adversarial perturbations to protect original audio recordings~\cite{sd97,sd98,sd96}. By introducing subtle perturbations or modifications at the data level, these techniques prevent a generative model from successfully cloning the voice during its inference procedure, i.e., the final cloned audio is dissimilar to the desired target audio. Zhang et al.~\cite{sd99} proposed a protection technique called \textit{POP} which applies imperceptible error-minimizing noises on original speech samples to prevent them from being effectively learned by TTS synthesis models. In their follow-up work, they proposed \textit{SPEC}~\cite{safespeech}, which they claim is robust against advanced adaptive adversaries. Dong et al.~\cite{sd10} proposed a \textit{Generative Adversarial Network (GAN)} based perturbation generation technique to protect against audio deepfake generation. Lastly, techniques such as \textit{AntiFake}~\cite{antifake}
use adversarial examples to disrupt the vocal timbre in synthesized audio, thereby thwarting zero-shot voice cloning and voice conversion by making the output dissimilar to the original speaker.

\subsection{Adversarial Purification}
Because protective perturbations function similarly to adversarial noise added to the input of VC models, an important question is whether existing adversarial purification techniques can remove these perturbations and thereby bypass the protection. Adversarial purification methods operate directly on the input audio and aim to recover a clean signal by eliminating adversarial distortions before the audio reaches the downstream VC system.

Unlike adversarial training, which modifies or retrains the model itself, purification does not require access to or control over the target VC model. This makes purification an appealing strategy in our setting, where the protection mechanism must defend against external VC models without altering them. Moreover, purification methods can generalize to unseen or adaptive perturbations because they focus on restoring the input signal rather than relying on a fixed set of adversarial examples encountered during training.

Recent research shows that generative model-based denoising approaches, especially diffusion models, are highly effective at removing perturbations from audio~\cite{sd141,sd142}. The technique by Wu et al.\cite{audiopure}, for instance, involves initially adding slight noise to the adversarial audio, then using a \emph{DDPM}\cite{Ho2020DenoisingDP} based diffusion model to denoise it and restore the clean audio. \emph{Diffpure}\cite{Nie2022DiffusionMF} uses a plug-and-play diffusion module as a pre-precessing step to remove perturbations. Guo et al.\cite{wavepurifier} build upon this work by building a more robust solution focused on Automated Speech Recognition (ASR), as ASR models often focus on the low frequency components and neglect high frequency components. They accomplish this by using a hierarchical diffusion framework using a pre-trained guided diffusion model. Tan et al. \cite{dualpure} employ a one-shot unconditional Mel spectrogram diffusion model and work on both time and frequency domains to remove perturbations. First, Gaussian perturbations are added into time domain signal through interpolation methods, then the samples' Mel spectrograms are purified with a diffusion model. While prior works primarily targeted adversarial perturbations added to compromise ASR models, Fan et al.~\cite{de-antifake} propose a two-step method to remove protective perturbations whose goal is to make speech un-learnable by generative models. In the first step, the perturbed audio is purified using an unconditional pre-trained diffusion model. In the second step, an Ornstein–Uhlenbeck SDE–based refinement model~\cite{sokol2013interventionornsteinuhlenbecksdes}, guided by phoneme information, is employed to more closely align the speech audio with the original signal.

In the context of image data, Li et al.~\cite{li2025adbm} investigate the limited robustness of purification methods based on simple pre-trained diffusion models. Specifically, they evaluate DiffPure~\cite{Nie2022DiffusionMF} under gradient-based PGD attacks~\cite{Madry2017TowardsDL} and demonstrate that both clean accuracy and adversarial robustness are significantly degraded. Although several follow-up efforts~\cite{di1,di2,di3,di4,di5} have sought to improve the robustness of DiffPure, their practical applicability is constrained by the high computational cost of Monte Carlo sampling required by these approaches~\cite{Cohen2019CertifiedAR}. To address these limitations, Li et al. propose ADBM, an approach that constructs a reverse bridge from the diffused adversarial image data distribution to the clean data distribution. Their method assumes that adversarial noise $\epsilon_a$ is injected at the beginning of the forward diffusion process during training. The model is trained to predict the original noise $\epsilon$ together with a scaled version of the adversarial noise $\epsilon_a$, enabling the reverse diffusion process to effectively remove adversarial perturbations while denoising the input.
\section{Threat Model and Security Objectives}
\label{sec:threat_model}
\begin{figure}[t]
    \centering \includegraphics[width=0.5\textwidth]{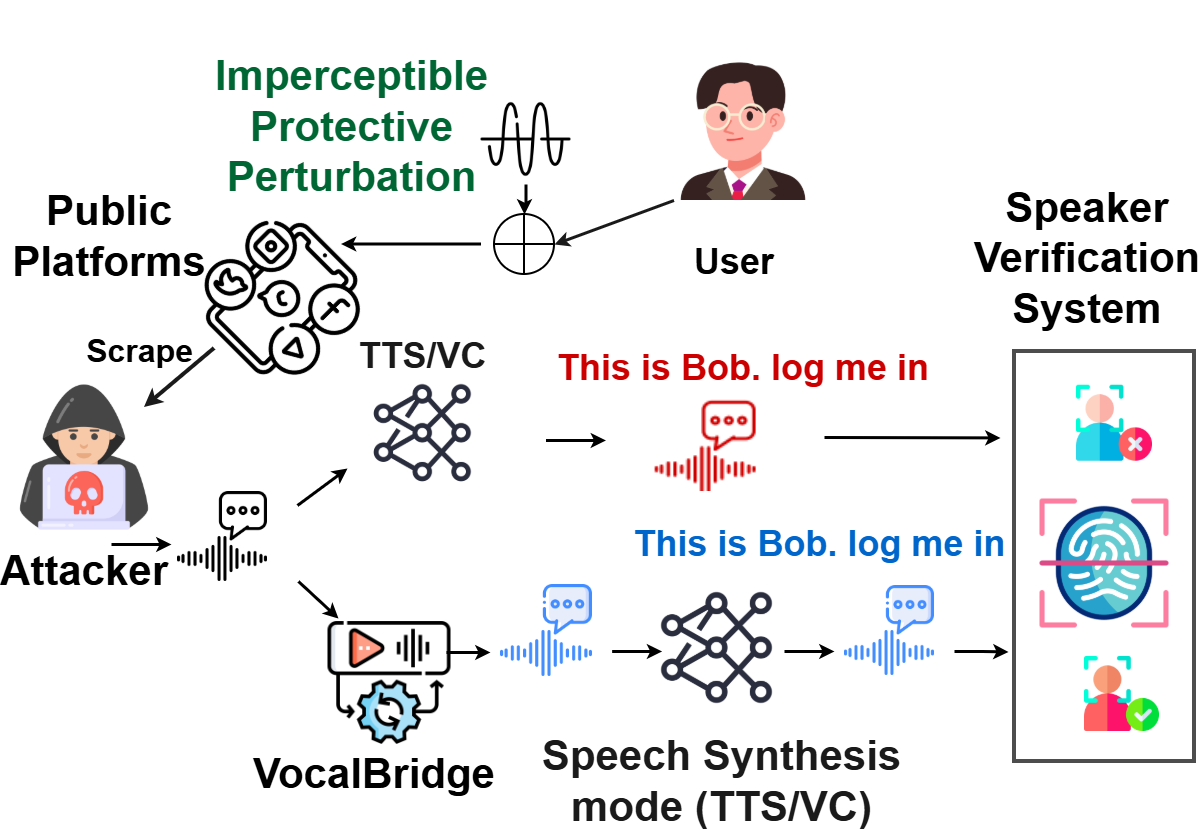}
    \caption{This figure illustrates the threat model, demonstrating how an attacker leverages VocalBridge to bypass existing defenses and execute voice-cloning attacks.}
    \label{fig:threatmodel}
\end{figure}

We consider the threat model shown in Fig.~\ref{fig:threatmodel}, where a \emph{user} seeks to prevent the misuse of their publicly available voice recordings, while an \emph{attacker} attempts to generate deepfake audio to impersonate the user and deceive ASV systems.
\\
\textbf{Parties and Objectives:}
The system consists of three components: the user, the adversary,   the ASV system.
\begin{itemize}
    \item \textbf{User.} The user legitimately produces utterances for communication or authentication and may publicly share them (e.g., on social media or online platforms). To prevent misuse, the user applies a defense mechanism $P_{\psi}$, parameterized by $\psi$, to each clean utterance $x$ before release, yielding a protected version:
    \[
    x_p = P_{\psi}(x).
    \]
    The goal of $P_{\psi}$ is to prevent malicious voice synthesis while preserving the naturalness and verifiability of the utterance for legitimate ASV use.
    \item \textbf{Attacker.} The Attacker collects publicly available protected utterances $\{x_p\}$ and attempts to clone or synthesize utterances that mimic the user’s timbre. The adversary’s objective is to generate a cloned utterance $\tilde{x}$ that is accepted by the ASV system as genuine.
    \item \textbf{Verifier (ASV System).} The ASV system evaluates a similarity score $s(\cdot,\cdot)$ between an input utterance and the enrollment utterance $x_e$ of the legitimate user. Verification is successful if $s(\cdot,\cdot) > \tau$, where $\tau$ is the system’s decision threshold.
\end{itemize}

\noindent \textbf{ User Objectives:} The defense mechanism $P_{\psi}$ should ensure that the released protected utterances $\{x_p\}$ maintain usability while resisting impersonation. Specifically, it should satisfy the following properties:
\begin{itemize}
    \item \textbf{Imperceptibility:} Perturbations introduced by $P_{\psi}$ must be imperceptible to human listeners so that $x_p$ remains natural, intelligible, and high-quality while preserving the linguistic content and intent of $x$.
    \item \textbf{Verifiability:} The legitimate user should still be accepted by the ASV system after protection:
    \[
    s(x_p, x_e) \geq \tau,
    \]
    where $s(\cdot,\cdot)$ is the cosine similarity between speaker embeddings, and $\tau$ is the ASV decision threshold.
    \item \textbf{Inimitability:} Synthetic (or deepfake) utterances $\tilde{x}$ generated from the protected samples $x_p$ should not be accepted by ASV as genuine:
    \[
    s(\tilde{x}, x_e) < \tau.
    \]
\end{itemize}

\noindent \textbf{Attacker Objectives:}
In our setting, the adversary’s goal is to \emph{invalidate the user’s protection objectives} through purification and synthesis. Specifically, the attacker aims to:
\begin{itemize}
    \item \textbf{Restore Verifiability:} 
   Purification aims to restore ASV verification by converting protected utterances that are initially rejected 
($s(x_p, x_e) < \tau$) into purified utterances that meet the acceptance threshold ($s(x_{\text{pur}}, x_e) \ge \tau$).

    \item \textbf{Preserve Imperceptibility:}
    Ensure that the purified audio remains natural and intelligible to human listeners.
    \item \textbf{Break Inimitability:}
    Enable synthetic or cloned speech generated from purified utterances to be accepted by ASV as genuine, thereby demonstrating that the user’s inimitability defense no longer holds.
\end{itemize}
\noindent Together, these objectives quantify the attack’s ability to reverse the user’s protective effects (restoring authentication functionality) while maintaining perceptual quality and facilitating re-cloning.

\noindent \textbf{Attacker Capabilities:}
The adversary operates under realistic and bounded capabilities:
\begin{itemize}
    \item The adversary can collect short segments of publicly available protected utterances $\{x_p\}$ and use them to train or prompt few-shot or zero-shot TTS/VC models.
    \item To improve synthesis, the adversary may apply a perturbation-removal mechanism $R_{\phi}$ (parameterized by $\phi$) to approximate the clean utterance:
    \[
    x_r = R_{\phi}(x_p).
    \]
    The purified utterance $x_r$ is then input to a generative model $G_{\theta}$ (parameterized by $\theta$) to produce the imitation $\tilde{x} = G_{\theta}(x_r)$.
    \item The adversary can access a small auxiliary dataset of paired samples $\{(x^{(i)}, x_p^{(i)})\}_{i=1}^N$ from non-target speakers to adapt or train $R_{\phi}$, but these pairs do not include the target user.
    \item The adversary has no white-box access to $P_{\psi}$ (i.e., no access to its parameters, gradients, or internals) and no knowledge of the internal parameters or thresholds of the ASV system.
\end{itemize}
\section{Audio Diffusion Bridge Model}
To align with the clean data distribution, we propose the \emph{Audio Diffusion-Bridge Model}, named \ourname , a purification mechanism that recovers clean audio from data protected by perturbation-based defenses by constructing a reverse bridge directly from the perturbed data distribution to the clean data distribution.

We build on the \emph{Adversarial Diffusion-Bridge Model (ADBM)} proposed by Li et al.~\cite{li2025adbm}, which was originally developed for image data. Directly applying ADBM to the audio domain is non-trivial due to the much higher temporal resolution, different noise characteristics, and the instability of waveform-level diffusion. To address these challenges, we adapt the ADBM formulation to speech and implement all diffusion operations within the latent space of a neural audio codec (EnCodec).This design allows diffusion to be performed on a more compact representation and enables the model to learn the mapping between protected and clean audio using paired training samples, rather than relying on classifier guidance as in the original ADBM. The decoder subsequently reconstructs the audio from the purified latent representation.
\begin{figure*}[t]
    \centering
    \includegraphics[width=0.8\textwidth]{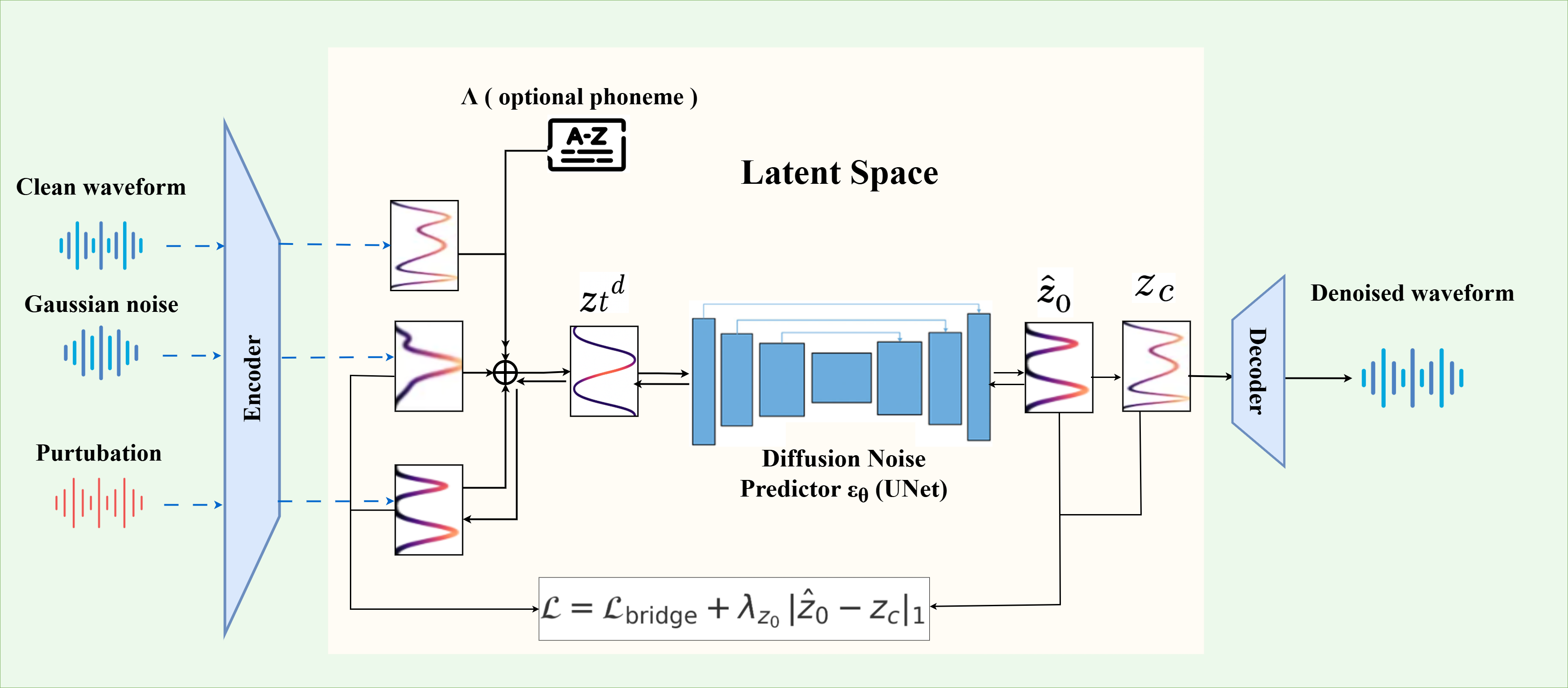}
    \caption{\ourname Training}
    \label{fig:VocalBridge_Training}
\end{figure*}

\subsection{Training Objective}
\label{sec:training_objective}

The model is a diffusion-based purification framework designed to suppress
\emph{ protective perturbation} in audio latent representations. Its forward process
follows the DDPM formulation, except that the initial state is perturbed by a
protection noise term $\varepsilon_a$ rather than beginning from the clean latent
sample. For each clean latent vector $\mathbf{z}_c$, the initial noised state is
defined as
\begin{equation*}
z_0^{a} = \mathbf{z}_c + \varepsilon_a.
\label{eq:z0_def}
\end{equation*}

The forward diffusion process is expressed as
\begin{equation*}
z_t^{a} = \sqrt{\bar{\alpha}_t}\, z_0^{a}
+ \sqrt{1 - \bar{\alpha}_t}\, \varepsilon,
\qquad
\varepsilon \sim \mathcal{N}(0,I), \quad 0 \le t \le T,
\label{eq:forward_process}
\end{equation*}
where $\bar{\alpha}_t$ denotes the cosine noise schedule and $T$ represents the
terminal diffusion step used for purification.

The reverse process learns a sequence $\{\hat{z}_t\}_{t:T \rightarrow 0}$ that maps
the diffused, protection-noised latent distribution $(z_T^{a})$ back to the clean
distribution $(\mathbf{z}_c)$. To align the intermediate diffusion trajectory,
scaling coefficients $c_{\text{in}}(t)$ and $c_{\text{tgt}}(t)$ are defined following
the derivation in~\cite{li2025adbm}:
\begin{equation*}
c_{\text{in}}(t)
=
\frac{\bar{\alpha}_T (1-\bar{\alpha}_t)}
{\sqrt{\bar{\alpha}_t}(1-\bar{\alpha}_T)},
\qquad
c_{\text{tgt}}(t)
=
\frac{\bar{\alpha}_T\sqrt{1-\bar{\alpha}_t}}
{(1-\bar{\alpha}_T)\sqrt{\bar{\alpha}_t}}.
\label{eq:c_in_c_tgt}
\end{equation*}

The bridged latent variable and the corresponding effective noise target are then
given by
\begin{align}
z_t^{d} &= \sqrt{\bar{\alpha}_t}\,\mathbf{z}_c
+ \sqrt{1-\bar{\alpha}_t}\,\varepsilon
+ c_{\text{in}}(t)\,\varepsilon_a,
\label{eq:bridge_zt} \\
\varepsilon_{\text{eff}} &= \varepsilon + c_{\text{tgt}}(t)\,\varepsilon_a.
\label{eq:bridge_eps}
\end{align}

A neural network $\varepsilon_\theta(z_t^{d},t)$ is trained to estimate the
effective noise term $\varepsilon_{\text{eff}}$. The \emph{bridge loss} is defined as
\begin{equation*}
\mathcal{L}_{\text{bridge}}
=
\mathbb{E}_{t,\varepsilon}
\Big[
\|\varepsilon_\theta(z_t^{d},t) - \varepsilon_{\text{eff}}\|_2^2
\Big].
\label{eq:bridge_loss}
\end{equation*}
This formulation is equivalent to the simplified bridge loss of
Li~\emph{et~al.}~\cite{li2025adbm}, up to constant scaling factors omitted for
clarity following Ho~\emph{et~al.}~\cite{Ho2020DenoisingDP}.

To enhance latent-space consistency, an auxiliary $L_1$ regularization term is
introduced using the reconstructed clean latent estimate:
\begin{equation*}
\hat{z}_0 =
\frac{
z_t^{d}
- \sqrt{1-\bar{\alpha}_t}\,\varepsilon_\theta(z_t^{d},t)
- \big(c_{\text{in}}(t)-\sqrt{1-\bar{\alpha}_t}\,c_{\text{tgt}}(t)\big)\varepsilon_a
}{\sqrt{\bar{\alpha}_t}}.
\label{eq:x0_estimate}
\end{equation*}
The total objective becomes
\begin{equation}
\mathcal{L}
=
\mathcal{L}_{\text{bridge}}
+ \lambda_{z_0}\,\|\hat{z}_0 - \mathbf{z}_c\|_1,
\qquad
\lambda_{z_0} \ll 1.
\label{eq:total_loss}
\end{equation}

As training progresses, the relative contribution of $\varepsilon_a$ decreases
with smaller $t$, enabling the model to progressively suppress protective noise
and recover the clean latent representation. Fig.~\ref{fig:VocalBridge_Training} illustrates the overall training process of \emph{\ourname}.

\subsection{Inference}
\label{sec:inference}

During inference, a protected waveform is first encoded into the latent
domain, producing the protected latent representation $\mathbf{z}_a$.
To initiate the reverse diffusion trajectory, we construct a noisy latent
state at terminal time $T$ as
\begin{equation*}
z_T = 
\sqrt{\bar{\alpha}_T}\,\mathbf{z}_a
\;+\;
\sqrt{1-\bar{\alpha}_T}\,\varepsilon,
\qquad
\varepsilon \sim \mathcal{N}(0,I).
\label{eq:inference_init}
\end{equation*}

A multi-step DDIM-style reverse process is then applied to map
$z_T \rightarrow \hat{z}_0$. At each diffusion step $t$, the denoiser
$\hat{\varepsilon}_\theta(z_t, t)$ predicts the noise present in the current
latent, and the clean latent estimate is obtained via
\begin{equation*}
\hat{z}_0 =
\frac{
z_t
-
\sqrt{1-\bar{\alpha}_t}\,
\hat{\varepsilon}_\theta(z_t, t)
}{
\sqrt{\bar{\alpha}_t}
}.
\label{eq:inference_ddim}
\end{equation*}

Iterating this update over a schedule
$t = T, t_{T-1}, \ldots, 0$ yields a final latent estimate
$\hat{z}_0$ that approximates the clean latent representation
$\mathbf{z}_c$. The purified waveform is obtained by decoding the
recovered latent through the EnCodec decoder.

\subsection{Whisper-Guided Phoneme Conditioning}
\label{sec:whisper_guided}

We extend our diffusion-bridge purifier with a lightweight phoneme-guided
conditioning mechanism that provides additional temporal structure for speech
denoising and refinement. A Whisper-based alignment front-end is used to
estimate approximate phoneme timings directly from the waveform, enabling the
model to better preserve speech content while aligning the purified output with
the clean speech distribution. Unlike prior works
(Fan \emph{et al.}~\cite{de-antifake}), which rely on ground-truth transcripts, our approach operates purely on
acoustic inputs and does not require transcript supervision.

Let $y$ denote the \emph{phoneme guidance signal}, a time-aligned sequence that
encodes the phoneme structure of an utterance. This signal is derived from a
phoneme alignment map $\Lambda$ produced by the Whisper-based aligner and
combined with a simple acoustic prior to form a smooth guidance track. During
training, $y$ is extracted from the clean waveform, standardized and passed
through a bounded nonlinearity, then RMS-matched to the bridged latent
$z_t^{d}$ before channel concatenation:
\[
x_{\text{in}} = \big[z_t^{d} \;\|\; \mathrm{RMSMatch}(y,\, z_t^{d};\, \gamma)\big].
\]
The denoising network is thus conditioned on both the noisy latent and the
phoneme guidance, enabling the reverse process to remove protective
perturbations while preserving speech structure. 

At inference time, the phoneme guidance signal $y$ is extracted directly from
the input audio using the same Whisper-based alignment module, without access
to clean references or transcripts. The resulting guidance acts as a soft
temporal prior rather than a strict constraint, making the method robust to
moderate alignment errors and residual noise. If phoneme alignment is
unavailable, the system automatically falls back to an unconditioned mode by
substituting a zero-valued guidance channel, ensuring stable and fully
compatible purification.

\subsection{Network Architecture}

The purifier network adopts a 1D U-Net architecture with time-step conditioning.
It operates entirely in the EnCodec latent domain, taking as input the bridged
latent $z_t^d$ (and optionally the phoneme guidance track $y$) and predicting the
effective noise term $\hat{\varepsilon}_\theta(z_t^d, t, y)$. The U-Net employs a
hierarchical encoder–decoder structure with residual Time-Delay Neural Network (TDNN) blocks to capture both
local and long-range temporal dependencies in the audio representation. Time
conditioning is implemented via sinusoidal embeddings and Feature-wise Linear Modulation (FiLM) modulation at each
decoder stage, enabling accurate diffusion-step awareness. 

In the phoneme-guided variant, an additional input channel concatenates the
normalized guidance signal $\Lambda(t)$ to the latent tensor, allowing linguistic
information to influence denoising while preserving the bridge consistency. The
network is trained using an AdamW optimizer with cosine learning rate decay and
gradient clipping for stability.

\section{Experiment and Evaluation}
We evaluate the proposed purification framework in terms of Restoration of Verifiability and perceptual quality, which together aim to break the defense mechanism’s inimitability property by enabling cloning of the protected speech and the generation of high-quality fake speech.

\label{sec:evaluation}
\subsection{Experimental Setup}
All experiments are implemented in PyTorch and conducted on an NVIDIA A100 server (80\,GB GPU, 512\,GB RAM) for model training and purification.
For lighter inference workloads, including TTS and VC synthesis, we additionally utilize NVIDIA Tesla~V100 GPUs (32\,GB).
The remaining experimental configurations are described in the following subsections.
\subsubsection{Datasets}
We conducted our experiments on two widely used benchmark speech datasets: \textbf{LibriSpeech}~\cite{librispeech} and \textbf{VCTK}~\cite{vctk2017}.  
LibriSpeech contains high-quality audiobook recordings from multiple speakers reading diverse transcripts, while VCTK consists of 110 English speakers with varied accents and recording conditions. For LibriSpeech, we select 40 speakers and use their first-chapter recordings, which provide phonetically balanced sentences suitable for evaluating intelligibility and speaker consistency. For VCTK, we use the first 50 utterances from each speaker. To train and evaluate our proposed purifier \ourname, we split the VCTK corpus into 30 speakers for training and 80 speakers for testing, and the LibriSpeech corpus into 27 speakers for training and 13 speakers for testing. All splits are gender-balanced to ensure a fair comparison across datasets.
In total, the evaluation uses 4,526 test samples and 13,414 training samples across both datasets. 

To construct evaluation data, we apply  protective-noise defenses to both datasets and then synthesize \emph{cloned} audio using selected TTS and VC tools.  
For TTS cloning, we sample previously unseen sentences from VCTK and assign each as the cloning target for a randomly chosen protected speaker.  
For VC cloning, we randomly select 50 utterances from source speakers and convert each into the voice of every protected speaker.  
This process produces approximately 42,289 synthesized samples per defense using TTS tools (209,936 total across all defenses) and 14,438 samples per defense using three VC tools (80,404 total VC samples). 

To evaluate \ourname 
Against baseline purification methods, each system is applied to the protected samples before cloning through both TTS and VC pipelines. For each purification method evaluated, we generate 4,526 purified test samples and a total of 290,340 synthesized samples (209,936 from TTS and 80,404 from VC). These synthesized outputs are then used to evaluate Restoration of Verifiability as well as perceptual quality.

\subsubsection{Protection Tools}
As discussed previously, perturbation-based voice defense methods aim to protect speech data by injecting optimized perturbations that make it unlearnable for TTS and VC models. In this work, we select a representative set of state-of-the-art defenses based on two criteria: 
(1) \textbf{Effectiveness:} each method is explicitly designed to degrade the ability of TTS and VC systems to synthesize realistic voices while preserving perceptual speech quality (\textit{quality defense}) and preserving speaker identity for ASV (\textit{timbre defense}). 
(2) \textbf{Availability:} We require that each defense method provides publicly released model checkpoints along with the corresponding optimization code.
 Consequently, we included publicly available perturbation-based speech defense methods that demonstrate strong reported performance. Specifically, we chose the following defenses in this
work:
\begin{itemize}
\item \textbf{SafeSpeech:} SafeSpeech~\cite{safespeech} proposes a training-time defense that targets both zero-shot and fine-tuning voice cloning attacks by making audio samples unusable for learning in TTS models. Its central mechanism, Speech PErturbative Concealment (SPEC), uses a surrogate generative model to guide the creation of perturbations that, as the authors claim, ensure there is nothing to learn from the protected data during model training. Rather than relying on inference-time adversarial examples, SafeSpeech introduces perturbations optimized through Mel-spectrogram loss and KL divergence with noise distributions, aiming to degrade both the speaker identity (timbre) and synthesis quality in any speech generated from the protected audio.

 \item \textbf{Attack-VC:} Attack-VC~\cite{attackvc} defends speeches by leveraging an encoder-decoder structure, where the encoder is divided into a content encoder and a speaker encoder. The content encoder, which captures the linguistic content, is left untouched to preserve what is being said, while the defense specifically targets the speaker encoder that extracts the unique identity of the speaker. The mechanism adds carefully crafted perturbations to the input utterances. These perturbations are designed to alter the output spectrogram, the speaker embedding, or both, so that even when the linguistic content remains clear, voice conversion models cannot accurately clone or identify the speaker’s voice. The changes are minimally perceptible to humans, but significantly reduce the risk of voice imitation or misuse by advanced VC technologies.The embedding attack offers the best trade-off between effectiveness and speed and is robust across models; accordingly, we use the embedding attack in our evaluation.
\item \textbf{Pivotal Objective Perturbation (POP):} POP~\cite{sd99} is a method that adds small imperceptible adversarial perturbations to audio prior to release, with the goal of making the data unlearnable for TTS voice cloning models. The authors claim that POP generates these perturbations by optimizing only the reconstruction loss (the difference between real and synthesized audio) and that, since this loss is shared across nearly all TTS models, the approach is universally effective, efficient, and transferable across architectures. They report that POP generates protected audio that sounds natural to human listeners, but when TTS models are trained on this protected data, the resulting synthetic speech becomes noisy and unusable.
\item \textbf{Anti-Fake:} AntiFake~\cite{antifake} targets identity disruption in synthesized speech by perturbing the speaker embedding space, ensuring that generated voices no longer resemble the original speaker to humans or machines. It employs two optimization strategies: a threshold-based method that pushes embeddings beyond a set distance from the original, and a target-based method that moves embeddings toward a different speaker. To ensure transferability to unknown TTS models, it optimizes perturbations using an ensemble of diverse speaker encoders. It also introduces a perceptual loss based on human hearing sensitivity and signal-to-noise ratios to preserve audio quality. Finally, it incorporates a human-in-the-loop process for selecting target voices and validating perceptual dissimilarity.  
 
\item \textbf{Active Defense Against Voice Conversion
Through Generative Adversarial Network (GAN-ADV): } GAN-ADV~\cite{advgan} introduces an adversarial defense framework that generates perturbations in the Mel-spectrogram domain using a generator-discriminator architecture (GAN), aiming to disrupt VC systems without perceptibly altering audio quality. The system includes a simulation module (SWCSM) that mimics the lossy process of waveform reconstruction and re-extraction of features, improving robustness to real-world inference pipelines. A substitute VC model is used during training to provide gradient signals, and perturbation optimization balances three objectives: fooling a discriminator (GAN loss), disrupting VC output (defense loss), and preserving audio fidelity (quality loss). Inference requires only the trained generator, making it efficient at deployment time.

\end{itemize}

We selected these public defenses because they demonstrate high performance and collectively cover methods that apply perturbations in different feature spaces and exhibit varying degrees of transferability across synthesis models. 
\textbf{SafeSpeech (SPEC)} introduces training-time waveform perturbations guided by a surrogate generative model. 
\textbf{Attack-VC} perturbs spectrogram representations either by directly distorting the output spectrogram (end-to-end), altering the speaker-embedding representation, or combining both through a feedback mechanism; among these, the \textit{speaker-embedding attack} variant is reported as the most effective, which we adopt. 
\textbf{POP} also operates in the waveform domain but optimizes perturbations with respect to Mel-spectrogram reconstruction loss. 
\textbf{AntiFake} manipulates speaker embeddings via threshold- and target-based optimization over an ensemble of encoders, enabling partial transferability across synthesis systems. 
\textbf{GAN-ADV} generates perturbations in the Mel-spectrogram domain through a GAN framework designed to remain robust after waveform reconstruction.

\subsubsection{Voice Cloning tools}
\label{a-tts-vc}
We assume that the adversary has access to a diverse set of state-of-the-art open-source TTS and VC models for synthesizing/cloning voice, each representing distinct architectural paradigms and recent advances in neural speech synthesis. We select six representative models 
\begin{itemize}[leftmargin=*, nosep]
    \item \textbf{VALL-E-X:} VALL-E-X~\cite{vallex} is a neural codec language model that predicts discrete acoustic tokens conditioned on source-language speech and target-language text. It supports zero-shot, cross-lingual TTS and speech-to-speech translation while preserving speaker timbre, emotion, and acoustic environment cues.
    
    \item \textbf{Tortoise-TTS:} Tortoise-TTS~\cite{tortoisetts} combines an autoregressive transformer-based acoustic model with a diffusion decoder and UnivNet vocoder. It first predicts compressed acoustic tokens and then refines them into expressive high-quality waveforms.
    
    \item \textbf{StyleTTS2:} StyleTTS2~\cite{styletts2} is a text-to-speech model that employs style diffusion and adversarial training with \emph{ large speech language models (SLMs)} to generate natural-sounding audio. It models speaking style as a latent random variable via diffusion, enabling style-consistent synthesis without requiring reference speech.
    
    \item \textbf{VQMIVC:} VQMIVC~\cite{vqmivc} is a one-shot voice conversion model that leverages vector quantization for content encoding and mutual information maximization for disentangling content, speaker, and pitch representations in an unsupervised manner.
    
    \item \textbf{HierSpeech++:} HierSpeech++~\cite{hierspeech} is a  hierarchical variational autoencoder integrating text-to-semantic-unit modeling, prosody control, and speech super-resolution. It incorporates normalizing flow modules for high-fidelity, zero-shot synthesis and conversion.
    
    \item \textbf{DiffHierVC~\cite{diffhiervc}:} A hierarchical VC framework utilizing two diffusion models, \textit{DiffPitch} and \textit{DiffVoice}, for sequential conversion. It achieves voice style transfer via a source–filter encoder that disentangles speech representations, with masked priors that enhance speaker adaptation quality.
\end{itemize}

\subsubsection{Speaker Verification Systems}
\label{sec:sv_systems}
We employ three standard ASV  
systems to evaluate speaker
identity preservation and recovery: \emph{$x$-vector, ECAPA, and $d$-vector}, implemented
using pretrained models from SpeechBrain~\cite{speechbrain2021} and
Resemblyzer~\cite{resemblyzer2020}. Each system maps an input waveform to a
fixed-dimensional embedding that represents the vocal identity of the speaker.
Verification between a test utterance and an enrolled speaker centroid is
performed via cosine similarity:
\begin{equation*}
\mathrm{score}(\tilde{\mathbf{e}}, \mathbf{c}) =
\frac{\tilde{\mathbf{e}}^{\top}\mathbf{c}}
     {\|\tilde{\mathbf{e}}\|_2\,\|\mathbf{c}\|_2},
\end{equation*}
where $\tilde{\mathbf{e}}$ denotes the test embedding and $\mathbf{c}$ is the
centroid of the speaker’s enrollment embeddings, computed as the average embedding over the enrollment utterances.

The decision threshold is determined using the
\emph{equal error rate (EER)} criterion, defined as the operating point where
the \emph{false-accept rate (FAR)} equals the \emph{false-reject rate (FRR)}:
\begin{equation*}
\mathrm{EER} = \mathrm{FAR}(\tau_{\mathrm{eer}})
             = \mathrm{FRR}(\tau_{\mathrm{eer}}),
\end{equation*}
with $\tau_{\mathrm{eer}}$ denoting the equal-error threshold.
All ASV embeddings and decision thresholds are computed using our evaluation dataset. To prevent data leakage, the subset used for threshold calibration (the development set) is strictly disjoint from the utterances used for enrollment and purification evaluation. A few clean utterances per speaker are used to compute speaker centroids, while the remaining utterances and their protected or purified counterparts serve as test trials. The ASV systems remain frozen throughout all experiments and function solely as objective evaluators of speaker identity consistency

On the development set 
which includes 149 speakers from our evaluation dataset, the x-vector, ECAPA, and d-vector systems achieve EERs of 0.0486, 0.006, and 0.0297, respectively, with corresponding thresholds of $\tau_{\mathrm{eer}}{=}0.951$ (x-vector), $0.419$ (ECAPA), and $0.750$ (d-vector). These calibrated thresholds are fixed for all subsequent purification and recovery experiments.

\subsubsection{Evaluation Metrics}
\label{sec:eval_metrics}
To assess the effectiveness of \ourname, we evaluate it with respect to the attacker objectives defined in Section~\ref{sec:threat_model}.
\\
\textbf{ $\bullet$ Restoration of Verifiability:}
Following the AntiFake evaluation design~\cite{antifake}, which measures
defense effectiveness through the \emph{Authentication Evasion Reduction Rate}
(AERR), we adopt the inverse perspective suitable for purification. Our goal is
to remove protective perturbations that suppress speaker verification and thus
\emph{restore authentication}. To this end, we introduce the
\emph{Authentication Restoration Rate (ARR)} as the primary metric to
quantify the efficacy of purification. ARR measures the proportion of previously
unverified (below-threshold) protected utterances that become successfully verified
after purification. For each identity, a clean enrollment centroid $\mathbf{c}$ is
calculated from a subset of clean enrollment utterances. Let $\mathrm{s}(\mathbf{e}, \mathbf{c})$
denote the cosine similarity between an embedding $\mathbf{e}$ and the enrolled centroid.
We denote by $s_i^{\mathrm{prot}}=\mathrm{s}(\mathbf{e}_i^{\mathrm{prot}},\mathbf{c})$ and
$s_i^{\mathrm{pur}}=\mathrm{s}(\mathbf{e}_i^{\mathrm{pur}},\mathbf{c})$ the similarity scores
for the protected and purified versions of the same utterance, respectively.
Given the equal-error threshold
$\tau_{\mathrm{eer}}$ in Section~\ref{sec:sv_systems}, the ARR is defined as
\begin{equation*}
\mathrm{ARR}(\tau_{\mathrm{eer}})=
\frac{\sum_i \mathbf{1}\{s_i^{\mathrm{prot}}<\tau_{\mathrm{eer}}\ \wedge\
s_i^{\mathrm{pur}}\ge\tau_{\mathrm{eer}}\}}
     {\sum_i \mathbf{1}\{s_i^{\mathrm{prot}}<\tau_{\mathrm{eer}}\}},
\label{eq:arr}
\end{equation*}
Higher ARR values indicate stronger authentication recovery and more effective
removal of protective perturbation. 
\\
\textbf{ $\bullet $ Imperceptibility:}
To evaluate perceptual fidelity and naturalness, we use the
objective \emph{Mean Opinion Score} (MOS) predicted by the
NISQA model~\cite{mittag2021nisqa}. NISQA estimates speech
quality and intelligibility on a scale from 1~(poor) to~5~(excellent),
providing an automated approximation of human perceptual
judgment. Higher MOS values indicate more natural and
intelligible speech, with scores above~3 generally reflecting good
audio quality\cite{antifake}.

We also use the Word Error Rate (WER)\cite{jiang2023mega} to assess pronunciation clarity. A pre-trained Whisper-small model [46] is employed for transcription due to its computational efficiency on our large dataset. Higher WER values indicate reduced speech clarity.

 
 \subsubsection{Purification Baselines}
 In this section, we provide a brief description of the denoising tools used in our experiments. The set includes adversarial denoisers and protective perturbations removers that represent a diverse range of recent approaches.
\begin{itemize}
\item \textbf{De-Antifake:} 
De-AntiFake\cite{de-antifake} is a voice cloning attack evaluation and purification system designed to test and defeat existing speech protection mechanisms that rely on adversarial perturbations. It simulates a realistic attacker who applies purification techniques to remove protective noise from speech before performing voice cloning. The tool introduces a new purification framework called \emph{PhonePuRe}, which works in two stages; (1) Purification Stage : Uses a diffusion-based model to clean adversarially perturbed audio. (2) Refinement Stage : Employs phoneme-guided alignment to fine-tune the purified speech so it closely matches natural, unprotected speech.
\item \textbf{WavePurifier:}
WavePurifier\cite{wavepurifier} is a defensive tool designed to purify audio adversarial examples that target ASR systems. It uses a hierarchical diffusion model that removes perturbations (via forward diffusion) and restores clean speech (via reverse diffusion). The tool divides spectrograms into frequency bands (low, mid, high) and optimizes purification intensity per band, maintaining speech quality while removing attacks. Evaluated on multiple ASR models and attacks, WavePurifier outperforms seven existing defenses, achieving the lowest Character and WERs and the highest purification success rate across diverse scenarios.
 \item \textbf{AudioPure:} AudioPure\cite{audiopure} is an adversarial purification-based defense pipeline made for acoustic systems using off-the-shelf diffusion models. It uses diffusion models to generate noise, which is added to adversarial audio in a small amount. Then, a reverse sampling step is performed to purify the noisy audio and recover the clean audio. It is a plug-and-play method, which can be applied to any pre-trained classifier without the need for additional fine-tuning or re-training. 
\item \textbf{DualPure:} DualPure\cite{dualpure} is a real-time purification based defense method against adversarial perturbations. First, it first disrupts the potential malicious perturbations at waveform level in the samples. Following this, an unconditional diffusion model is used to purify the features at the frequency level. Specifically, it first applies a \emph{time-domain purifier (TDP)} to purify waveform signals, then converts the waveform to a mel spectrogram and applies frequency-domain purification. It achieves good adversarial robustness against both white-box and black-box attacks.
\end{itemize}

\subsection{Experiment Results}

\begin{table*}[t!]
\centering
\scriptsize
\setlength{\tabcolsep}{2.5pt}   
\renewcommand{\arraystretch}{1.1}
\caption{Authentication Restoration Rate (\%) for selected TTS models.}
\resizebox{\textwidth}{!}{%
\begin{tabular}{| l l | ccc | ccc | ccc | ccc | ccc | ccc|}
\hline
\multirow{2}{*}{\textbf{Protection}} &
\multirow{2}{*}{\textbf{TTS models}} &
\multicolumn{3}{c|}{\textbf{De-AntiFake}} &
\multicolumn{3}{c|}{\textbf{DualPure}} &
\multicolumn{3}{c|}{\textbf{WavePurifier}} &
\multicolumn{3}{c|}{\textbf{AudioPure}} &
\multicolumn{3}{c|}{\textbf{VocalBridge}} &
\multicolumn{3}{c}{\textbf{VocalBridge-W}} \\
\cline{3-20}
& & xvec & ecapa & dvec & xvec & ecapa & dvec & xvec & ecapa & dvec & xvec & ecapa & dvec & xvec & ecapa & dvec & xvec & ecapa & dvec \\
\hline
\textbf{AntiFake}
& StyleTTS2    & 86.61 & 40.34 & 54.05 & 13.11 & 0.40 & 0.94 & 52.02 & 8.72 & 1.40 & 19.97 & 1.54 & 1.65 & 58.17 & 21.74 & 2.76 & 55.27 & 12.71 & 7.24 \\
& Tortoise-TTS & 62.05 & 53.75 & 59.90 & 19.57 & 1.11 & 0.98 & 24.37 & 9.66 & 0.84 & 25.56 & 4.55 & 2.07 & 47.86 & 35.86 & 8.37 & 42.83 & 30.34 & 11.57 \\
& VALL-E-X     & 45.42 & 26.20 & 25.36 & 63.72 & 7.94 & 1.12 & 25.79 & 6.61 & 0.37 & 59.41 & 3.44 & 1.94 & 38.13 & 14.68 & 11.54 & 48.91 & 16.43 & 16.44 \\

\rowcolor{blue!10}
\emph{Avg (TTS)}
& & 64.69 & 40.10 & 46.44
& 32.13 & 3.15 & 1.01
& 34.06 & 8.33 & 0.87
& 34.98 & 3.18 & 1.89
& 48.05 & 24.09 & 7.56
& 49.00 & 19.83 & 11.75 \\

\hline
\textbf{Attack-VC}& StyleTTS2     & 32.32 & 10.12 & 9.07  & 48.54 & 36.93 & 54.47 & 5.29  & 14.30 & 10.05 & 45.27 & 29.51 & 47.32 & 61.21 & 60.03 & 67.23 & 63.86 & 60.66 & 64.37 \\
                  & Tortoise-TTS  & 32.52 & 9.60  & 3.94  & 36.56 & 23.89 & 28.29 & 3.45  & 16.26 & 2.53  & 41.74 & 19.90 & 20.52 & 38.07 & 25.91 & 15.67 & 38.30 & 24.40 & 10.03 \\
                  & VALL-E-X      & 29.47 & 8.00  & 3.36  & 73.03 & 52.84 & 25.25 & 15.63 & 12.24 & 1.93  & 69.51 & 42.71 & 19.02 & 39.67 & 36.02 & 19.79 & 41.48 & 35.51 & 24.97 \\
\rowcolor{blue!10} \emph{Avg (TTS)} &          & 31.44 & 9.24  & 5.46  & 52.71 & 37.89 & 36.00 & 8.12  & 14.27 & 4.84  & 52.18 & 30.71 & 28.95 & 46.32 & 40.66 & 34.23 & 47.88 & 40.19 & 33.12 \\
\hline
\textbf{GAN-ADV}  & StyleTTS2     & 39.82 & 14.93 & 6.09  & 17.37 & 2.44  & 4.05  & 6.24  & 7.56  & 1.87  & 14.41 & 2.43  & 5.04  & 49.01 & 36.92 & 31.31 & 57.58 & 40.24 & 34.22 \\
                  & Tortoise-TTS  & 32.76 & 10.88 & 2.64  & 28.05 & 5.99  & 3.11  & 3.71  & 8.19  & 0.68  & 22.99 & 7.26  & 2.62  & 38.99 & 16.79 & 9.90  & 49.11 & 12.04 & 7.25  \\
                  & VALL-E-X      & 29.80 & 8.92  & 3.27  & 58.56 & 8.68  & 4.73  & 12.94 & 4.67  & 0.95  & 61.95 & 5.94  & 4.69  & 40.86 & 19.98 & 17.71 & 47.99 & 20.83 & 18.45 \\
\rowcolor{blue!10} \emph{Avg (TTS)} &          & 34.13 & 11.58 & 4.00  & 34.66 & 5.70  & 3.96  & 7.63  & 6.81  & 1.17  & 33.12 & 5.21  & 4.12  & 42.95 & 24.56 & 19.64 & 51.56 & 24.37 & 19.97 \\
\hline
\textbf{POP}      & StyleTTS2     & 38.91 & 30.21 & 43.38 & 14.24 & 0.00  & 0.00  & 10.82 & 14.77 & 5.82  & 12.35 & 0.09  & 0.00  & 40.00 & 50.04 & 58.76 & 45.49 & 52.78 & 60.50 \\
                  & Tortoise-TTS  & 33.75 & 17.92 & 12.81 & 28.10 & 2.66  & 2.44  & 2.69  & 7.39  & 0.43  & 22.02 & 1.25  & 1.30  & 42.86 & 15.09 & 11.75 & 45.06 & 18.04 & 10.94 \\
                  & VALL-E-X      & 31.02 & 20.98 & 17.77 & 59.92 & 7.59  & 7.19  & 8.49  & 10.12 & 1.11  & 54.92 & 6.59  & 5.11  & 35.16 & 31.03 & 40.87 & 35.26 & 29.25 & 43.35 \\
\rowcolor{blue!10} \emph{Avg (TTS)} &          & 34.56 & 23.04 & 24.66 & 34.09 & 3.42  & 3.21  & 7.33  & 10.76 & 2.45  & 29.76 & 2.64  & 2.14  & 39.34 & 32.05 & 37.13 & 41.94 & 33.36 & 38.27 \\
\hline
\textbf{SafeSpeech}& StyleTTS2    & 58.75 & 21.67 & 17.89 & 33.48 & 2.04  & 0.90  & 20.28 & 7.85  & 1.97  & 33.72 & 1.56  & 2.09  & 54.87 & 21.13 & 31.01 & 56.44 & 22.92 & 31.06 \\
                  & Tortoise-TTS  & 49.30 & 20.54 & 13.38 & 34.66 & 5.04  & 3.84  & 14.21 & 7.79  & 1.04  & 28.83 & 7.49  & 4.00  & 39.44 & 5.88  & 5.25  & 33.91 & 4.90  & 5.78  \\
                  & VALL-E-X      & 35.29 & 21.50 & 7.65  & 62.49 & 3.55  & 3.93  & 10.97 & 3.53  & 0.22  & 63.71 & 4.28  & 1.98  & 41.65 & 15.33 & 16.18 & 37.03 & 12.67 & 17.42 \\
\rowcolor{blue!10} \emph{Avg (TTS)} &          & 47.78 & 21.24 & 12.98 & 43.54 & 3.54  & 2.89  & 15.15 & 6.39  & 1.08  & 42.09 & 4.44  & 2.69  & 45.32 & 14.11 & 17.48 & 42.46 & 13.50 & 18.09 \\
\hline
\end{tabular}%
}
\label{tab:table-ARR-TTS}
\end{table*}

\begin{table*}[ht!]
\centering
\scriptsize
\setlength{\tabcolsep}{2.5pt}
\renewcommand{\arraystretch}{1.1}
\caption{Authentication Restoration
Rate (\%) for selected VC models.}
\resizebox{\textwidth}{!}{%
\begin{tabular}{| l l | ccc | ccc | ccc | ccc | ccc | ccc |}
\hline
\multirow{2}{*}{\textbf{Protection}} &
\multirow{2}{*}{\textbf{VC models}} &
\multicolumn{3}{c|}{\textbf{De-AntiFake}} &
\multicolumn{3}{c|}{\textbf{DualPure}} &
\multicolumn{3}{c|}{\textbf{WavePurifier}} &
\multicolumn{3}{c|}{\textbf{AudioPure}} &
\multicolumn{3}{c|}{\textbf{VocalBridge}} &
\multicolumn{3}{c}{\textbf{VocalBridge-W}} \\
\cline{3-20}
& & xvec & ecapa & dvec & xvec & ecapa & dvec & xvec & ecapa & dvec & xvec & ecapa & dvec & xvec & ecapa & dvec & xvec & ecapa & dvec \\
\hline
\textbf{AntiFake}
& DiffHierVC   & 77.39 & 23.99 & 74.05 & 8.70 & 1.54 & 1.42 & 40.69 & 2.85 & 4.82 & 18.12 & 1.54 & 1.42 & 37.05 & 5.57 & 13.42 & 26.81 & 3.73 & 9.96 \\
& HierSpeechpp & 2.75  & 2.89  & 16.75 & 1.61 & 8.64 & 31.28 & 0.00 & 6.94 & 17.77 & 10.92 & 29.28 & 37.63 & 2.40 & 14.93 & 45.69 & 35.54 & 51.93 & 75.63 \\
& VQMIVC       & 6.32  & 17.84 & 33.03 & 0.00 & 0.34 & 0.00 & 0.24 & 0.00 & 0.00 & 0.00 & 0.00 & 0.00 & 67.65 & 37.97 & 5.83  & 72.30 & 43.73 & 8.52  \\

\rowcolor{blue!10}
\emph{Avg (VC)}
& & 28.82 & 14.91 & 41.27
& 3.44 & 3.51 & 10.90
& 13.64 & 3.26 & 7.20
& 9.01 & 10.27 & 13.68
& 35.70 & 19.49 & 21.65
& 44.88 & 33.13 & 31.37 \\

\hline
\textbf{Attack-VC} & DiffHierVC   & 13.79 &  2.28 & 11.34 & 13.79 & 36.48 & 50.85 & 46.67 & 13.08 & 22.03 & 22.58 & 18.77 & 43.38 & 30.00 & 54.18 & 73.45 & 34.38 & 53.21 & 70.34 \\
                  & HierSpeechpp &  1.49 &  0.78 &  8.15 &  3.36 & 20.10 & 33.03 &  0.00 &  5.75 & 10.70 &  6.29 & 30.60 & 41.28 &  4.67 & 24.75 & 42.51 & 19.72 & 41.63 & 56.88 \\
                  & VQMIVC       & 48.85 & 35.59 & 27.93 & 38.04 & 41.70 & 55.68 & 33.36 &  7.98 &  2.70 & 35.59 & 39.00 & 38.92 & 58.66 & 48.21 & 58.38 & 59.29 & 51.29 & 58.92 \\
\rowcolor{blue!10}\emph{Avg (VC)} &              & 21.38 & 12.88 & 15.81 & 18.40 & 32.76 & 46.52 & 26.68 &  8.94 & 11.81 & 21.49 & 29.46 & 41.19 & 31.11 & 42.38 & 58.11 & 37.80 & 48.71 & 62.05 \\
\hline
\textbf{GAN-ADV}  & DiffHierVC    & 25.91 &  3.60 &  2.66 & 13.64 &  1.46 &  3.98 &  8.65 &  2.57 &  4.31 & 13.84 &  1.51 &  5.27 & 22.07 & 11.95 & 25.27 & 35.14 & 17.43 & 38.34 \\
                  & HierSpeechpp  & 17.83 &  2.77 &  2.22 &  3.55 &  0.82 &  2.77 &  1.40 &  1.21 &  4.10 &  9.66 &  1.35 &  2.36 & 10.24 &  7.99 & 17.09 & 33.80 & 14.53 & 29.59 \\
                  & VQMIVC        & 20.06 & 25.19 & 15.86 & 23.33 &  8.39 & 15.65 & 14.39 &  2.11 &  2.04 & 24.74 &  6.21 &  6.80 & 24.67 & 25.00 & 17.69 & 59.39 & 43.17 & 30.61 \\
\rowcolor{blue!10}\emph{Avg (VC)} &               & 21.27 & 10.52 &  6.91 & 13.51 &  3.56 &  7.47 &  8.15 &  1.96 &  3.48 & 16.08 &  3.02 &  4.81 & 19.00 & 14.98 & 20.02 & 42.78 & 25.04 & 32.85 \\
\hline
\textbf{POP}      & DiffHierVC    &  6.25 &  5.09 & 36.13 &  5.26 &  0.47 &  0.27 & 60.00 &  9.29 &  2.73 &  0.00 &  0.69 &  0.27 & 10.00 & 22.38 & 28.42 & 14.29 & 25.93 & 36.61 \\
                  & HierSpeechpp  &  0.83 &  2.00 & 29.86 &  0.00 &  0.00 &  0.00 &  0.74 &  0.54 &  1.10 &  0.79 &  0.27 &  0.00 &  0.76 &  7.26 & 17.63 & 29.01 & 38.07 & 49.59 \\
                  & VQMIVC        & 21.43 & 32.17 &  7.94 & 27.52 & 15.38 & 19.30 & 10.89 &  5.26 &  0.00 & 21.43 & 14.29 &  7.94 & 37.61 & 29.33 & 28.07 & 38.74 & 35.06 & 33.33 \\
\rowcolor{blue!10}\emph{Avg (VC)} &              &  9.50 & 13.09 & 24.64 & 10.93 &  5.28 &  6.52 & 23.88 &  5.03 &  1.28 &  7.41 &  5.08 &  2.74 & 16.13 & 19.66 & 24.71 & 27.35 & 33.02 & 39.84 \\
\hline
\textbf{SafeSpeech} & DiffHierVC   & 46.13 &  8.81 & 24.97 & 23.69 &  0.84 &  0.62 & 31.56 &  3.33 &  9.38 & 19.57 &  1.82 &  4.20 & 36.96 &  5.14 & 27.34 & 39.14 &  6.71 & 34.94 \\
                    & HierSpeechpp & 61.44 & 11.95 & 38.10 & 23.65 &  1.40 &  1.40 & 34.23 &  3.97 & 11.92 & 15.24 &  2.50 &  5.63 & 44.76 &  8.97 & 36.21 & 46.15 & 10.44 & 39.28 \\
                    & VQMIVC       & 29.62 & 15.09 & 29.47 & 26.39 & 11.84 & 16.10 & 23.24 &  7.56 & 12.68 & 30.89 & 11.52 & 13.62 & 63.01 & 32.47 & 16.59 & 66.95 & 35.94 & 19.02 \\
\rowcolor{blue!10}\emph{Avg (VC)} &  & 45.73 & 11.95 & 30.85 & 24.58 &  4.69 &  6.04 & 29.68 &  4.95 & 11.33 & 21.90 &  5.28 &  7.82 & 48.24 & 15.53 & 26.71 & 50.75 & 17.70 & 31.08 \\
\hline
\end{tabular}%
}
\label{tab:table-ARR-VC}
\end{table*}

\begin{figure}[t]
    \centering
    \includegraphics[width=\columnwidth]{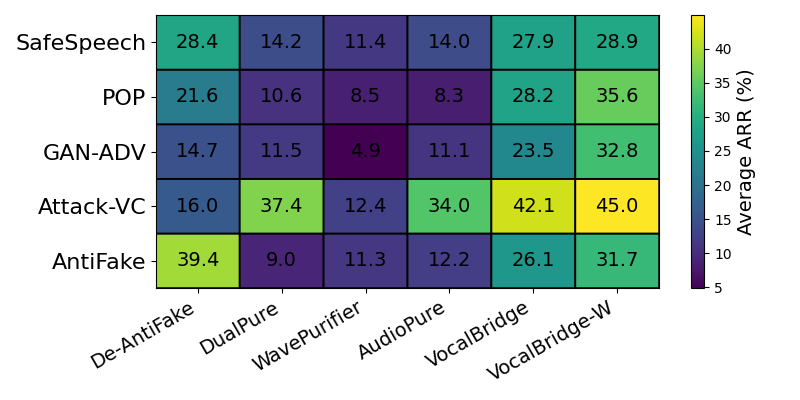}
    \caption{The mean ARR (\%) over three speaker verification back-ends (x-vector, ECAPA-TDNN, and d-vector).}
    \label{fig:All}
\end{figure}

\subsubsection{Restoring Verifiability: \ourname{} vs. Baselines}
In this section, we evaluate VocalBridge against four purification mechanisms that an attacker can employ: De-AntiFake (full model including Purification and Refinement), AudioPure, DualPure, and WavePurifier. Among these, De-AntiFake is the only related work explicitly designed to defeat voice-cloning defenses by removing their protective perturbations; to the best of our knowledge, no other purification-based attack has been proposed for this setting. The remaining purifiers were originally developed to mitigate adversarial attacks on ASR systems and are not tailored to our protection-removal threat model. Nevertheless, we include them as baselines to determine whether these  purification techniques can inadvertently act as effective protection-removal attacks.
 Table~\ref{tab:table-ARR-TTS} and Table~\ref{tab:table-ARR-VC} report the ARR of speech synthesized by selected TTS and VC models from purified datasets across the three ASV systems: x-vector, ECAPA-TDNN, and d-vector.
As summarized in Fig.~\ref{fig:All}, VocalBridge-W achieves the strongest overall restoration across protections. Under the Attack-VC protection, VocalBridge-W reaches 45.0\% ARR, improving over the best existing method (DualPure at 37.4\%) by a margin of 7.6. On the GAN-ADV and POP protections, VocalBridge-W attains 32.8\% and 35.6\%, exceeding the strongest baselines by 18.1 and 14.0, respectively. On SafeSpeech, VocalBridge-W also provides the highest restoration (28.9\%), slightly surpassing prior work. VocalBridge exhibits similar gains, raising Attack-VC performance from 37.4\% to 42.1\%, GAN-ADV from 14.7\% to 23.5\%, and POP from 21.6\% to 28.2\%. The only setting where our models do not lead is AntiFake, where the specialized De-AntiFake method remains higher (39.4\% vs.\ 31.7\% for VocalBridge-W). Notably, De-AntiFake requires full access to the AntiFake detector and its ground-truth transcript to operate its refiner, a requirement rarely satisfied in practice, whereas VocalBridge and VocalBridge-W function without any privileged model access, making them substantially more practical and broadly applicable.



\begin{figure*}[t]
    \centering
    \includegraphics[width=0.8 \textwidth]{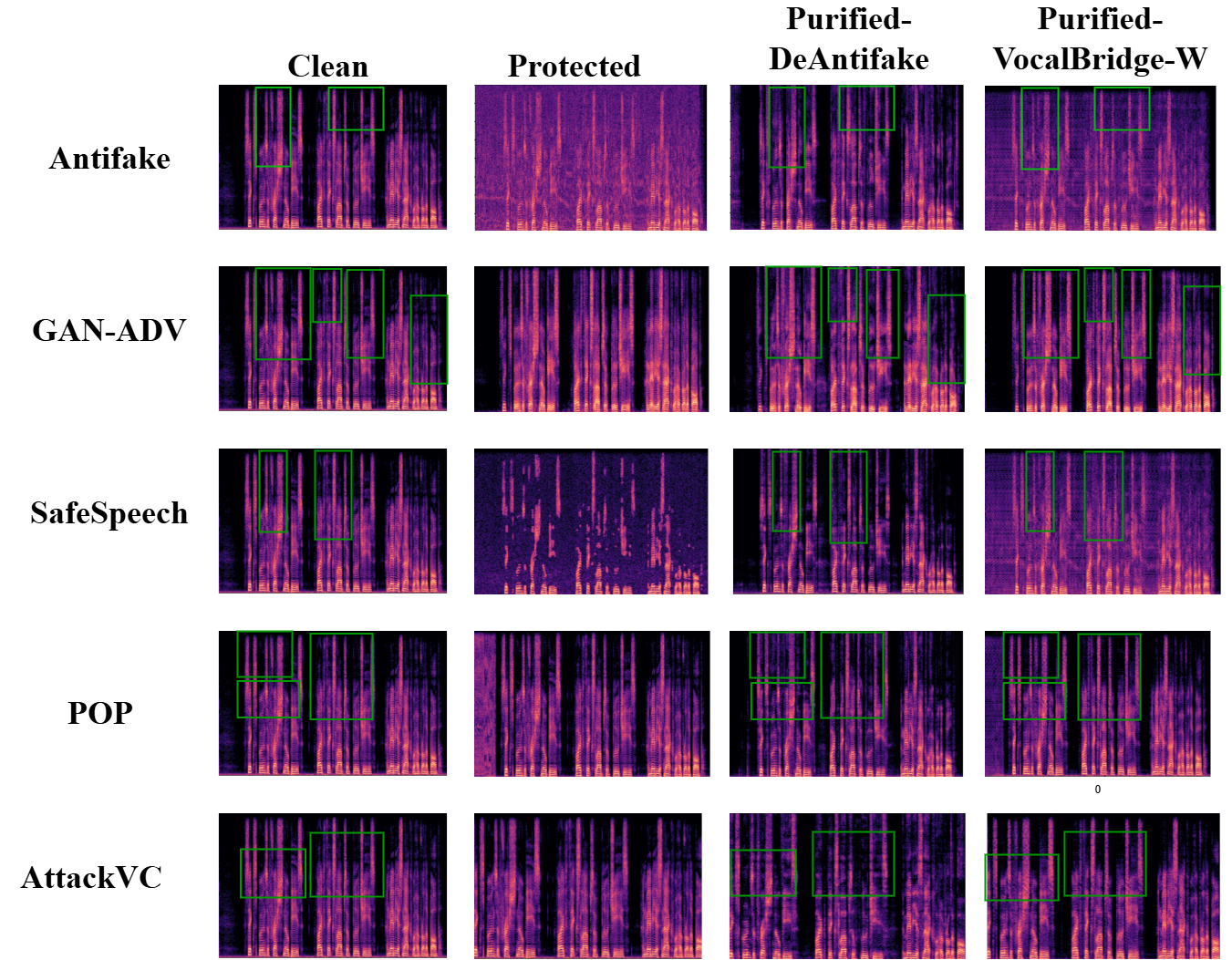}
    \caption{The spectrogram comparison shows that \ourname-W better maintains the structure and detail of the clean samples than the leading baseline De-AntiFake.}
    \label{fig:p232_pop_clean_spec}
\end{figure*}

Fig.~\ref{fig:p232_pop_clean_spec} shows the spectrograms of the input and output samples for voice cloning. The visual comparison indicates that \ourname-W preserves the structural and spectral details of the clean audio more effectively than the state-of-the-art De-AntiFake baseline. A notable limitation of De-AntiFake is its reliance on clean transcripts, an assumption that often breaks under noisy or distorted conditions such as SafeSpeech and AntiFake. In contrast, VocalBridge and VocalBridge-W require no ground truth transcripts and therefore maintain effectiveness even when transcripts are degraded or unavailable, providing a more practical basis for robustness evaluation.

Overall, these results indicate that VocalBridge and VocalBridge-W provide more reliable purification than existing methods, effectively removing protective perturbations and restoring speaker identity across different protection mechanisms, ASV backends, and synthesis architectures.

\subsubsection{Evaluation of Cross-Perturbation Generalization of \ourname}
To assess the generalization capability of our purification framework, we use a \emph{Mono} model that is trained on only a \emph{single} perturbation pattern.  
We train \emph{Mono} using only the perturbation pattern produced under \emph{GAN-ADV} and then apply it to audio perturbed by other protections, without any form of adaptation. 
This setup allows us to evaluate whether learning from just one perturbation type is sufficient for removing a wide range of unseen perturbations. Table~\ref{tab:cross} shows that the Mono variant remains closely aligned with the Adaptive model across all protection types.

Overall, these results demonstrate that the \ourname maintains high authentication-restoration performance even on perturbation patterns it never encountered during training, indicating strong robustness and generalization across heterogeneous purification defenses.

\begin{table}[!ht]
\centering
\scriptsize
\setlength{\tabcolsep}{2pt}
\renewcommand{\arraystretch}{1.05}
\caption{Authentication Restoration Rate (\%) for VC and TTS under Adaptive and Mono purification models}
\resizebox{\columnwidth}{!}{%
\begin{tabular}{|l|l|ccc|ccc|}
\hline
\multirow{2}{*}{\textbf{Protection}} &
\multirow{2}{*}{\textbf{Model}} &
\multicolumn{3}{c|}{\textbf{\ourname}} &
\multicolumn{3}{c|}{\textbf{\ourname(Mono)}} \\
\cline{3-8}
& & xvec & ecapa & dvec & xvec & ecapa & dvec \\
\hline

& DiffHierVC     & 37.05 &  5.57 & 13.42 & 18.12 &  2.62 &  2.87 \\
& HierSpeechpp   &  2.40 & 14.93 & 45.69 &  2.40 & 13.72 & 32.49 \\
\textbf{AntiFake} & VQMIVC       & 67.65 & 37.97 &  5.83 & 66.90 & 36.61 &  3.59 \\
& StyleTTS2      & 58.17 & 21.74 &  2.76 & 33.13 &  4.83 &  2.15 \\
& Tortoise-TTS   & 47.86 & 35.86 &  8.37 & 38.78 &  6.43 &  2.22 \\
& VALL-E-X       & 38.13 & 14.68 & 11.54 & 34.00 & 12.46 & 12.04 \\
\rowcolor{blue!10}\emph{Avg (All)} 
&                & \textbf{41.88} & \textbf{21.79} & \textbf{14.60} & \textbf{32.22} & \textbf{12.78} & \textbf{9.23} \\
\hline

& DiffHierVC     & 10.00 & 22.38 & 28.42 & 15.00 & 28.47 & 30.60 \\
& HierSpeechpp   &  0.76 &  7.26 & 17.63 &  3.05 &  8.99 & 12.95 \\
\textbf{POP}& VQMIVC         & 37.61 & 29.33 & 28.07 & 42.48 & 32.47 & 31.58 \\
& StyleTTS2      & 40.00 & 50.04 & 58.76 & 49.17 & 49.68 & 53.05 \\
& Tortoise-TTS   & 42.86 & 15.09 & 11.75 & 42.33 & 13.45 &  7.69 \\
& VALL-E-X       & 35.16 & 31.03 & 40.87 & 34.86 & 32.07 & 37.61 \\
\rowcolor{blue!10}\emph{Avg (All)} 
&                & \textbf{27.73} & \textbf{25.86} & \textbf{30.92} & \textbf{31.15} & \textbf{27.52} & \textbf{28.91} \\
\hline

& DiffHierVC     & 30.00 & 54.18 & 73.45 & 30.00 & 54.55 & 63.84 \\
& HierSpeechpp   &  4.67 & 24.75 & 42.51 &  4.73 & 24.21 & 31.50 \\
\textbf{Attack-VC}& VQMIVC       & 58.66 & 48.21 & 58.38 & 57.84 & 55.98 & 52.43 \\
& StyleTTS2      & 61.21 & 60.03 & 67.23 & 61.31 & 59.97 & 63.85 \\
& Tortoise-TTS   & 38.07 & 25.91 & 15.67 & 38.61 & 26.49 & 13.34 \\
& VALL-E-X       & 39.67 & 36.02 & 19.79 & 39.82 & 35.48 & 19.43 \\
\rowcolor{blue!10}\emph{Avg (All)} 
&                & \textbf{38.71} & \textbf{41.52} & \textbf{46.17} & \textbf{38.72} & \textbf{42.78} & \textbf{40.73} \\
\hline

& DiffHierVC     & 36.96 &  5.14 & 27.34 & 28.96 &  4.14 & 14.48 \\
& HierSpeechpp   & 44.76 &  8.97 & 36.21 & 44.33 &  7.22 & 15.87 \\
\textbf{SafeSpeech} & VQMIVC     & 63.01 & 32.47 & 16.59 & 57.42 & 30.77 & 10.73 \\
& StyleTTS2      & 54.87 & 21.13 & 31.01 & 47.08 & 16.98 & 20.29 \\
& Tortoise-TTS   & 39.44 &  5.88 &  5.25 & 28.63 &  4.44 &  2.45 \\
& VALL-E-X       & 41.65 & 15.33 & 16.18 & 35.82 & 11.47 &  9.54 \\
\rowcolor{blue!10}\emph{Avg (All)} 
&                & \textbf{46.78} & \textbf{14.82} & \textbf{22.10} & \textbf{40.37} & \textbf{12.50} & \textbf{12.23} \\
\hline
\end{tabular}
}
\label{tab:cross}
\end{table}

\subsection{Effectiveness of Whisper-Guided Phoneme Conditioning}
 Integrating Whisper features into VocalBridge (forming \emph{VocalBridge-W in Tables~\ref{tab:table-ARR-TTS} and \ref{tab:table-ARR-VC}}) improves purification stability and helps the model preserve speaker-relevant phonetic structure. Across all protection settings, Whisper-guided conditioning provides consistent gains over our simple VocalBridge purifier. For instance, on GAN-ADV, VocalBridge-W improves encoder-averaged ARR of VC models from 19/15/20\% to 43/25/33\%, and on POP, ARR rises from 16/20/25\% to 27/33/40\%. 
\subsubsection{Evaluation of Speech Quality and Imperceptibility}
\begin{figure}[t]
    \centering
    \includegraphics[width=0.95\linewidth]{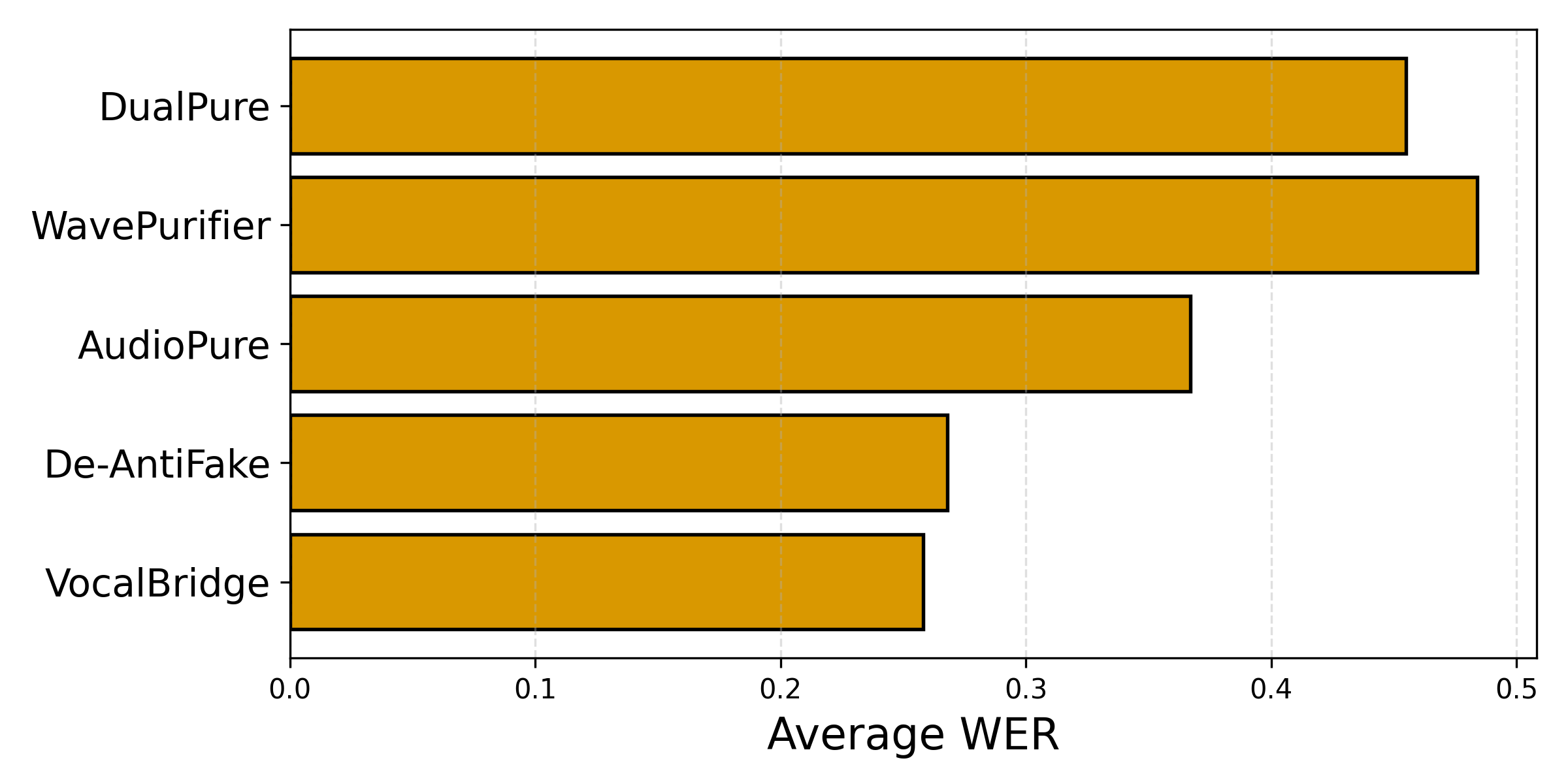}
    \caption{
        Average WER of protection mechanisms on synthetic voices. 
    }
    \label{fig:wer_comparison}
\end{figure}

Table \ref{t-mos} reports the average NISQA-TTS MOS across all VC/TTS systems. The protected speech retains perceptual quality close to the original (3.36 vs. 3.57). Existing purification defenses exhibit substantially lower quality, with average MOS values ranging from 2.95 to 3.27. Our method matches the quality of the protected samples (3.36) while outperforming all prior purification approaches. 
Fig.~\ref{fig:wer_comparison} reports the average WER for different protection mechanisms on synthetic speech. \ourname attains the lowest WER (0.258), outperforming all competing approaches. 

These results show that \ourname preserves high perceptual fidelity and avoids the over-smoothing and spectral distortions present in baseline methods, allowing attacker-used VC/TTS models to replicate speaker characteristics more accurately.
\begin{figure}[t]
    \centering
    \includegraphics[width=0.95\linewidth]{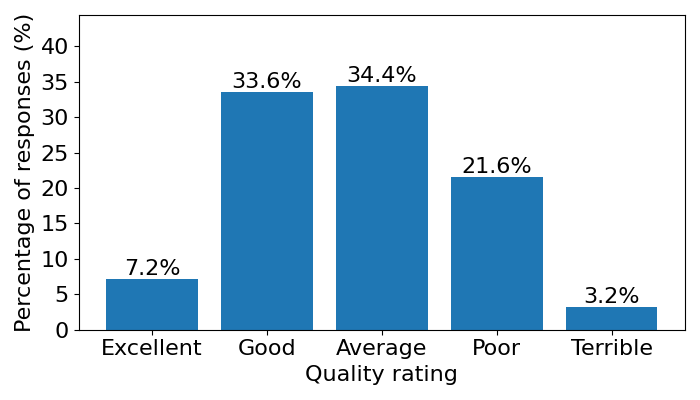}
    \caption{
        ARR under different adaptive strategies. }
    \label{fig:user}
\end{figure}
\subsection{Subjective Evaluation of Speech Quality}
We conducted a user study with 27 participants to evaluate the subjective quality of the audio generated by our system. The study consisted of a demographic questionnaire followed by a set of listening questions in which participants rated the perceptual quality of synthesized audio created by cloning purified samples with \ourname. Participants rated each audio clip using a four-point scale: Good, Average, Poor, or Terrible. The survey included audio obtained by purifying perturbations introduced by Antifake, Attack-VC, GAN-ADV, POP, and SafeSpeech, and subsequently cloning the purified audio with Tortoise-TTS; each participant was presented with a random subset of these samples. One attention-check question containing obvious white-noise corruption was included to ensure participant reliability. As shown in Fig.~\ref{fig:user}, 75.2\% of all ratings fell within the Good, Average, or Excellent categories, with only a small fraction marked as Poor or Terrible, indicating that the purified-and-cloned audio is generally perceived as natural and intelligible. The survey was created using Qualtrics and deployed through Amazon Mechanical Turk.
\begin{table*}[t]
\centering
\scriptsize
\setlength{\tabcolsep}{3pt}
\renewcommand{\arraystretch}{1.15}
\caption{Mean Opinion Score (MOS) predicted by NISQA-TTS (1–5, higher is better).}
\begin{tabular}{|l|c|c|c|c|c|c|c|}
\hline
\textbf{VC / TTS} &
\textbf{Original} &
\textbf{Protected} &
\textbf{De-AntiFake} &
\textbf{DualPure} &
\textbf{WavePurifier} &
\textbf{AudioPure} &
\textbf{\ourname (Ours)} \\
\hline
StyleTTS2     & 3.68 & 3.61 & 3.13 & 2.88 & 3.31 & 2.86 & 3.59 \\
Tortoise-TTS  & 3.92 & 3.58 & 3.44 & 3.34 & 2.40 & 3.23 & 3.62 \\
VALL-E-X      & 3.59 & 3.27 & 3.05 & 3.20 & 2.74 & 3.19 & 3.29 \\
DiffHierVC    & 3.43 & 3.28 & 3.47 & 2.93 & 3.25 & 2.86 & 3.24 \\
HierSpeechpp  & 3.88 & 3.63 & 3.81 & 3.41 & 3.54 & 3.23 & 3.59 \\
VQMIVC        & 2.91 & 2.78 & 2.73 & 2.82 & 2.46 & 2.80 & 2.81 \\
\rowcolor{blue!10} \emph{Avg (All)} & \textbf{3.57} & 3.36 & 3.27 & 3.10 & 2.95 & 3.03 & \textbf{3.36} \\
\hline
\end{tabular}
\label{t-mos}
\end{table*}

 \subsection{Adaptive Protection}
We also evaluate an adaptive protection scenario in which the protection mechanism has white-box access to our purification model, including its gradients, and can therefore optimize its perturbations accordingly.
 We adopt the adaptive-attack methodology used in De-Antifake\cite{de-antifake}. Because the overall purification function, from input waveform to purified output, is effectively non-differentiable due to components such as EnCodec quantization and stochastic diffusion sampling, we apply their Backward Pass Differentiable Approximation (BPDA) strategy. BPDA treats the purifier as an identity mapping during backpropagation, yielding surrogate gradients that enable end-to-end perturbation optimization despite these non-differentiable operations. As in De-Antifake, we also use Expectation Over Transformation (EOT) to account for randomness in the diffusion process. We average gradient estimates over 1, 5, 10, and 15 stochastic runs, which provide a stable gradient approximation under stochastic sampling.

\begin{figure}[t]
    \centering
    \includegraphics[width=0.95\linewidth]{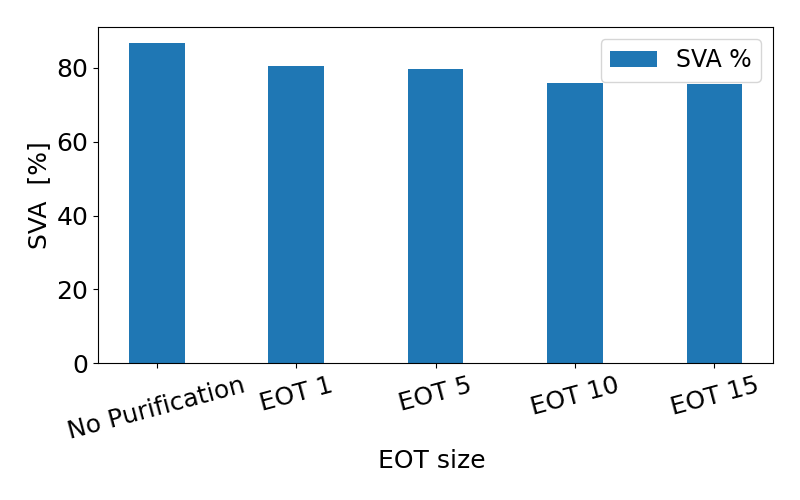}
    \caption{
        SVA under different adaptive strategies. }
    \label{fig:Adaptive_ASV}
\end{figure}

\begin{figure}[t]
    \centering
    \includegraphics[width=0.95\linewidth]{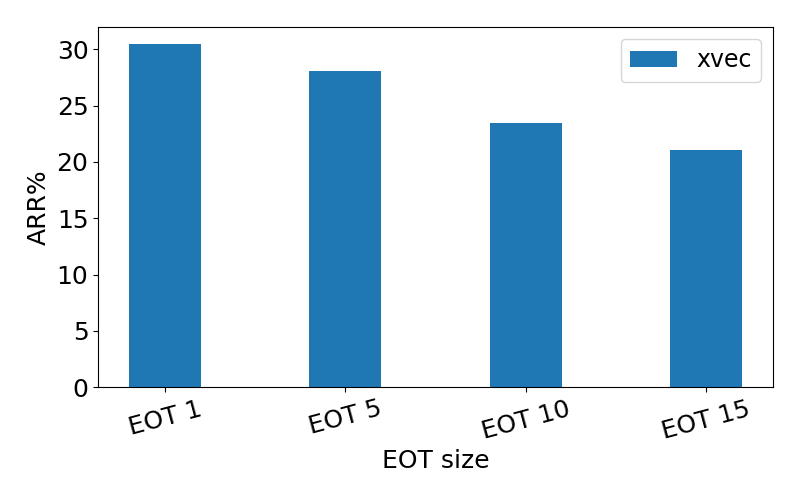}
    \caption{
        ARR under different adaptive strategies. }
    \label{fig:Adaptive_ARR}
\end{figure}

To evaluate the effectiveness of our method, we first purify the speech that has been adaptively protected, using \ourname-W, and then clone the resulting purified audio with StyleTTS2. We measure the attack success using  \emph{ARR} and \emph{Speaker Verification Accuracy (SVA)} with the x-vector model, where SVA is defined as the fraction of cloned samples accepted as genuine by the ASV system, computed as the average binary match decision across all cloned samples. As shown in Fig.~\ref{fig:Adaptive_ASV} and Fig.~\ref{fig:Adaptive_ARR}, under these adaptive strategy settings, ASV remains above 75\% and ARR stays above 20\%. These results indicate that, even with white-box knowledge, developing effective adaptive protections against our purification method is still difficult, highlighting the risks we have identified.
\section{Discussion and Limitations}
\subsection{Phoneme-Guided Refinement}
Our method uses the Whisper \textit{small} model to generate phoneme-level alignments for $\Lambda$-guided refinement. This choice was made because of limited computational resources, and it reduces alignment accuracy, especially when the audio is noisy or adversarial. As a result, the phoneme guidance provides only small improvements. The resource limitations also prevented us from experimenting with larger or more robust alignment models. Future work could explore stronger Whisper variants to obtain more reliable $\Lambda$ features and greater denoising benefits.

\subsection{Ethics Consideration}
The rapid advancement of speech synthesis technologies has ensued a race between voice cloning techniques and protective countermeasures. By highlighting the flaws in the existing perturbation-based countermeasures, this study adds to this adversarial dynamic. We recognize the work's multifaceted implications; while the primary motivation is to assess the robustness of existing safeguards and thus motivate the development of more resilient solutions, the proposed Diffusion-Bridge purification model also serves as an effective method for circumventing those same safeguards. 
We firmly believe that the potential benefits of enhancing safeguards against unauthorized speech synthesis and voice cloning far outweigh the risks of misuse of our findings, particularly as we intend to not make our source code publicly accessible. In the interest of open research and advancing science, we will grant access to the source code upon request and only after a thorough review of the request and confirmation of its intent. To ensure that the developers of the evaluated protection tools can appropriately respond and make necessary adaptations, we plan to disclose our findings to them upon acceptance of this manuscript and definitely prior to the publication of this work.

The results of our work highlight an urgent need for the community to rethink perturbation-based approaches and explore fundamentally new strategies for safeguarding voice data. Future protection mechanisms must be designed with robustness to advanced pre-processing and purification in mind, ensuring they remain effective as adversarial capabilities continue to advance.

\section{Conclusion}

This paper examined the vulnerability of contemporary voice-protection mechanisms to purification-based attacks and introduced VocalBridge, a bridged latent-diffusion model designed for reconstructing speaker identity. VocalBridge achieves effective identity recovery across both TTS and VC pipelines, and its Whisper-guided extension, VocalBridge-W, further stabilizes phonetic structure and improves reconstruction quality further. 

Our evaluations show that VocalBridge and VocalBridge-W substantially outperform existing purification approaches. We additionally assess the generalization capabilities of our model, demonstrating that attackers do not require explicit knowledge of a protection system’s perturbation pattern to mount successful reconstruction attacks. Our adaptive-protection study shows that even with complete white-box access, creating robust defenses against our purification method is still challenging, emphasizing the severity of the risks we expose.

Overall, our results reveal that current speech-protection techniques remain vulnerable to latent-space diffusion–based purification attacks. By releasing our evaluation framework and bridged-diffusion models, we aim to support the development of stronger and more principled defenses for speech privacy and synthetic-media authentication.


\bibliographystyle{IEEEtran}
\bibliography{References}
\appendices

\end{document}